\documentclass[12pt]{iopart}
\usepackage{iopams}  
\usepackage{graphicx}
\begin{document}

\title[Pre-Steady-State Morphogen Gradients]{When it Pays to Rush: Interpreting Morphogen Gradients Prior to Steady-State}

\author{Timothy Saunders and Martin Howard}
\address{
Department of Computational and Systems Biology,
John Innes Centre, 
Norwich,
NR4 7UH,
United Kingdom
}
\ead{martin.howard@bbsrc.ac.uk}
\begin{abstract}
During development, morphogen gradients precisely determine the position of gene expression boundaries despite the inevitable presence of fluctuations. Recent experiments suggest that some morphogen gradients may be interpreted prior to reaching steady-state. Theoretical work has predicted that such systems will be more robust to embryo-to-embryo fluctuations. By analysing two experimentally motivated models of morphogen gradient formation, we investigate the positional precision of gene expression boundaries determined by pre-steady-state morphogen gradients in the presence of embryo-to-embryo fluctuations, internal biochemical noise and variations in the timing of morphogen measurement. Morphogens that are direct transcription factors are found to be particularly sensitive to internal noise when interpreted prior to steady-state, disadvantaging early measurement, even in the presence of large embryo-to-embryo fluctuations. Morphogens interpreted by cell-surface receptors can be measured prior to steady-state without significant decrease in positional precision provided fluctuations in the timing of measurement are small. Applying our results to experiment, we predict that Bicoid, a transcription factor morphogen in {\it Drosophila}, is unlikely to be interpreted prior to reaching steady-state. We also predict that Activin in {\it Xenopus} and Nodal in zebrafish, morphogens interpreted by cell-surface receptors, can be decoded in pre-steady-state.

\end{abstract}

\vspace{2pc}
\maketitle

\section{Introduction}

A central paradigm for spatial positioning in embryonic development is the morphogen gradient: signaling molecules (morphogens) form a concentration gradient across the system with gene expression boundaries determined by whether the morphogen concentration is greater or less than a particular threshold \cite{Gurdon2001,Lander2007}. Despite the apparent simplicity of the morphogen concept there are still many unanswered questions including: how morphogens are transported through the embryo \cite{Lander2002,Gonzalez-Gaitan2003,Kruse2004,Spirov2009,Yu2009,Nahmad2009}; how the system measures the morphogen concentration \cite{Gurdon2001,Melen2005,Dessaud2007}; what mechanisms control the timing of morphogen interpretation \cite{Lander2007,Schier2005,Hagos2007,Tucker2008}; and how robust morphogen gradients are to fluctuations \cite{Barkai2007,Tostevin2007,Saunders2009}.

Experimental and theoretical approaches often assume that morphogen gradients are in steady-state prior to measurement \cite{Saunders2009,Kicheva2007,Eldar2003}. However, recent experimental evidence suggests that some morphogen gradients may be interpreted before reaching steady-state including Sonic Hedgehog (SHH) \cite{Dessaud2007,Harfe2004}, Hedgehog \cite{Nahmad2009}, Activin \cite{Gurdon1995}, Bone Morphogenetic Protein (BMP) \cite{Tucker2008,Mizutani2005}, Nodal \cite{Harvey2009} and, controversially, Bicoid (Bcd) \cite{Bergmann2007}. Theoretical analysis suggests that such pre-steady-state readout can provide reliable positioning of gene boundaries \cite{Mizutani2005,Saha2005} and it may even be preferable to steady-state measurement if there are large variations between embryos in the morphogen production rate \cite{Bergmann2007}. Furthermore, time-varying morphogen concentrations offer the additional advantage that they can define the expression of multiple genes at similar spatial positions but at different times \cite{Gurdon1995,Nahmad2009}. 

Morphogen gradients are known to provide very precise positional information about cell fate despite embryos being inherently noisy \cite{Houchmandzadeh2002,Gregor2007b,Bollenbach2008,He2008,Manu2009}. The gradient must therefore be robust to external embryo-to-embryo variations, such as differences in size \cite{Houchmandzadeh2002,He2008} and morphogen production rates \cite{Eldar2003}.  Internal fluctuations, caused by processes such as protein production, diffusion and degradation, are present in all biological systems and limit the precision of biochemical signaling \cite{Tostevin2007,Saunders2009,Berg1977}. To reduce internal fluctuations, the embryo time-averages the morphogen gradient \cite{Tostevin2007,Gregor2007b}. However, in pre-steady-state, the concentration varies systematically as a function of time and so variations in the onset of averaging (as well as in the averaging period) could reduce precision \cite{Bergmann2007}. For the first time, our work considers all the above fluctuations to build a more complete understanding of the precision of pre-steady-state morphogen gradient interpretation.

We analyse two experimentally relevant models of morphogen gradient formation \cite{Lipshitz2009}: morphogen production restricted to a localised region with global diffusion and degradation; and morphogen production by a spatially distributed mRNA gradient \cite{Spirov2009}. In the latter case, morphogen is produced throughout the embryo but is confined to the cell/nucleus in which it is produced. We also apply our analysis to a simplified model of the Bcd gap-gene regulatory network \cite{Bergmann2007,Jaeger2004a} to demonstrate that our approach extends to more realistic systems.  We find that internal fluctuations present a major barrier to precise pre-steady-state interpretation of morphogen gradients in both models. In particular, if the morphogen is a direct transcription factor (with nanometer-scale DNA binding sites on the target genes) then internal fluctuations can make pre-steady-state decoding especially imprecise. We apply our analysis to Bcd, which has recently provoked debate over whether it is interpreted prior to steady-state \cite{Cell-Argument2008a,Cell-Argument2008b}. We find that a Bcd diffusion rate of $1{\mu}m^2s^{-1}$ (as suggested by Bergmann et al. \cite{Bergmann2007,Cell-Argument2008a}) is inconsistent with precise position determination within the known time window of Bcd target gene expression due to error from internal fluctuations. 

Our results can be applied to morphogens interpreted by cell-surface receptors, such as Activin in {\it Xenopus} and Nodal in zebrafish. The effective measuring measuring volume is large for such morphogens (roughly the cell size) and hence the effects of internal fluctuations on the precision with which the morphogen is interpreted by the cell are small. Provided that fluctuations in the onset time of measurement are small, we predict that pre-steady-state measurement precision for such systems is not significantly different from that in steady-state, meaning pre-steady-state interpretation is possible. 

\section{Models}

We examine two models of morphogen gradient formation: localised morphogen \underline{s}ynthesis with long-range \underline{d}iffusion and spatially uniform linear \underline{d}egradation (SDD); and spatially-distributed \underline{m}\underline{R}NA \underline{g}radients (MRG) with no extra-cellular/extra-nuclear diffusion. The SDD model is the prevailing model for morphogen propagation \cite{Gregor2007a} and is representative of diffusion-dominated systems; for example, the Activin gradient in {\it Xenopus} forms through passive diffusion \cite{Gurdon1994}. However, recent experiments on Bcd in {\it Drosophila}, suggest that the {\it bcd} mRNA itself may form a spatially distributed gradient \cite{Spirov2009,Lipshitz2009}. The MRG model provides a simple description of such gradient formation. Our models encapsulate the properties of these mechanisms for morphogen gradient formation, enabling a more general discussion of the (dis)advantages of pre-steady-state interpretation.

{\bf SDD Model}: In steady-state the SDD model results in an exponentially decaying concentration profile. Bcd and Activin are observed to have such concentration profiles in steady-state \cite{Houchmandzadeh2002,McDowell1997}. The morphogen concentration $\rho_{sdd}(\mathbf{x},t)$ at position ${\mathbf{x}}$ and time $t$ is described by
\begin{equation}
\label{eq:Model1}
\partial_t{\rho}_{sdd}(\mathbf{x},t) = D\nabla^2{\rho}_{sdd}(\mathbf{x},t) - \mu{\rho}_{sdd}(\mathbf{x},t)\,,
\end{equation}
where $D$ is the diffusion constant and $\mu$ the rate of morphogen degradation. The morphogen is produced along the plane at $x=0$ with constant flux $J$, leading to the boundary condition $D\partial_x\rho(\mathbf{x},t)|_{x=0}+J=0$. In the absence of fluctuations, the general solution of (\ref{eq:Model1}) is \cite{Bergmann2007}
\begin{equation}
\label{eq:Solution1}
\rho_{sdd}(\mathbf{x},t) = \left(1-f(x,t)\right)\rho_{0}e^{-x/\lambda_{sdd}}\,,
\end{equation}
where $\lambda_{sdd}=\sqrt{D/\mu}$ is the decay length, $\rho_0=J\lambda_{sdd}/D$, $f(x,t) = \textrm{erfc}(z_{-})/2 + e^{2x/\lambda_{sdd}}\textrm{erfc}(z_{+})/2$, $\textrm{erfc}(z)$ is the complimentary error function, $z_{\mp} = (2Dt/\lambda_{sdd}\mp x)/\sqrt{4Dt}$, and $x,\lambda_{sdd}{\ll}L$ (where $L$ is the system length). In steady-state the morphogen concentration is $\rho_0e^{-x/\lambda_{sdd}}$. The morphogen concentration depends only on the time after initiation and the linear distance from the plane at $x=0$. Solving (\ref{eq:Model1}) using a geometry similar to that of the {\it Drosophila} embryo results in no significant difference in the morphogen profile from the approach described above \cite{Bergmann2007}, justifying our methodology.

{\bf MRG Model}: We assume the underlying mRNA gradient is fully formed before measurement begins, consistent with experimental observations \cite{Spirov2009,Lipshitz2009}. Including spatially uniform morphogen degradation, the morphogen concentration $\rho_{mrg}(\mathbf{x},t)$ is described by $\partial_t\rho_{mrg}(\mathbf{x},t) = j-\nu\rho_{mrg}$, where $j=j_0e^{-x/\lambda_{mrg}}$ is the morphogen production rate per unit volume and $\nu$ the rate of morphogen degradation. In the absence of fluctuations, the concentration profile is given by
\begin{equation}
\label{eq:Model2}
\rho_{mrg}(\mathbf{x},t) = \left(1-e^{-\nu{t}}\right)\rho_{1}e^{-x/\lambda_{mrg}}\,,
\end{equation}
where $\rho_{1}=j_{0}/\nu$. The gradient is always exponential with decay length $\lambda_{mrg}$ (independent of $\nu$), but with increasing amplitude as time increases. 

Figure~1A shows the evolution of (\ref{eq:Solution1}) and (\ref{eq:Model2}) with time at fixed position, $x_T$, using $\mu=\nu$ and $\rho_0=\rho_1$. $\rho_{mrg}$ increases more rapidly initially than $\rho_{sdd}$ as morphogen is produced locally at $x_T$. As morphogen begins to diffuse into the measuring region, $\rho_{sdd}$ rapidly increases. Finally, at large times the two models tend to the steady-state concentration. As we now demonstrate, the different evolution of the morphogen concentrations with time (Figure~1A inset) can have an important impact on the positional precision of the two models in the presence of different sources of noise. 

\section{Results}

\subsection{Internal Biochemical Fluctuations}
Morphogen production, diffusion and degradation are stochastic processes, resulting in internal fluctuations in the morphogen concentration about the threshold position. We are interested in the positional error, $\omega_{int}$, due to such fluctuations. However, before we calculate the spatial precision of the two models in the presence of internal fluctuations, we first consider when and how the system interprets the morphogen gradient. At early times in one possible scenario, morphogen interpretation could be suppressed by an inhibitory protein distributed uniformly across the system. The inhibitor could, for example, compete with the morphogen and its cofactor by binding cooperatively to morphogen transcription factor binding sites. When the inhibitor density decreases (for example, by degradation) below a threshold, repression will be released and morphogen interpretation switched on at that particular time throughout the system. Even after this switch on, the system cannot make instantaneous measurements of the morphogen gradient as the probability of a receptor/DNA binding site being occupied by a morphogen molecule is often small \cite{Tostevin2007,Gregor2007b,Dyson1998}. To increase precision, developmental systems sample the morphogen gradient over a period of time, referred to here as the `time-averaging window' (TAW). The period of the TAW, $\tau_m$, can be controlled by the lifetimes of the downstream products and feedback between the morphogen and these downstream products. For example, chick neural development is partly controlled by a gradient of SHH, but cells become gradually desensitized to ongoing SHH exposure due to SHH-dependent upregulation of a protein that inhibits SHH signaling \cite{Dessaud2007}.

During morphogen interpretation, the time between independent measurements of the morphogen concentration scales as $\tau_{ind}\sim({\Delta}x)^2/D_0$, where ${\Delta}x$ is the linear dimension of the measuring region \cite{Tostevin2007,Berg1977} and $D_0$ is the local diffusion constant in the vicinity of the receptor/DNA binding site. Note that while there is no extra-cellular/extra-nuclear diffusion in the MRG model, the morphogen can diffuse within the cell/nucleus.

In our simple model, the statistics of the particle number at a given position due to internal fluctuations are Poissonian \cite{Tostevin2007,Wu2007,Lepzelter2008}. We have confirmed this result in pre-steady-state using three-dimensional stochastic simulations for both the SDD and MRG models, Figure~1B. The stochastic simulations involved simulating the propagation of up to 3000 particles using Monte Carlo techniques and measuring the particle occupation at specified sites (see \cite{Tostevin2007} for details). Furthermore, we assume that $\tau_m\ll\tau$, meaning that the gradient varies comparatively little over the averaging period, which begins at time $\tau$. This assumption is reasonable for Bcd, where target gene expression begins around 90 minutes after initiation but the TAW is restricted by the relatively short time between successive nuclear divisions \cite{Gregor2007b}. Hence, we calculate $\omega_{int}$ using \cite{Tostevin2007}
\begin{equation}
\label{eq:omega0}
\omega_{int}\approx A\tau_m^{-1/2}\frac{\left\langle {\rho_{\tau,\tau_m}}(x_T)\right\rangle^{1/2}}{|\left\langle\partial_x{\rho}_{\tau,\tau_m}(x_T)\right\rangle|}\,,
\end{equation}
where $\langle...\rangle$ denotes ensemble averaging, $\rho_{\tau,\tau_m}(x)$ is the average morphogen concentration over the period $\tau\le{t}\le\tau_m$ (see Appendix A), and $A$ is a constant determined by details of the time-averaging network (see Appendix A). To test the validity of (\ref{eq:omega0}), we compared the positional error found from stochastic simulations of the SDD model on small systems (see \cite{Tostevin2007} for details) with (\ref{eq:omega0}). As can be seen from Figure~1C, the agreement is good. Four properties of (\ref{eq:omega0}) are particularly important: (i) longer averaging times result in reduced positional error from internal fluctuations; (ii) low concentration results in increased precision due to reduced fluctuations in the density; (iii) shallow gradients result in large positional error; (iv) the constant $A\propto(\Delta{x})^{-1/2}$ in three dimensions, see Appendix A, so smaller measuring volumes result in larger error due to internal fluctuations. Therefore, we expect morphogenetic transcription factors (${\Delta}x$ nanometre scale, the typical size of a DNA binding site) to have larger error than morphogens interpreted by cell-surface receptors ($\Delta{x}$ micrometre scale, the typical size of a cell). In the latter case, although the receptors themselves are small, the effective morphogen measuring volume is comparable to the cell size \cite{Berg1977,Bialek2005}.

At early times the morphogen gradient is flat, accentuating any concentration fluctuations into large positional error, see Figure~2A (Figures~2 and~3 use parameters consistent with the direct transcription factor Bcd \cite{Bergmann2007,Gregor2007b,Gregor2007a} unless stated otherwise). Early measurement in both the SDD and MRG models comes at a major cost to precision. The SDD model is particularly imprecise at short times, even including spatial averaging \cite{Gregor2007b}, as insufficient morphogen has diffused into the measuring region for reliable measurement. The MRG model is more precise than the SDD model at early times as morphogen is produced locally at the threshold position, resulting in a steeper morphogen profile, see Figure~2A. However, the precision is still worse than at steady-state. It is clear from Figure~2A that, given a particular final steady-state profile, improved precision is achieved through fast morphogen production, diffusion and degradation (i.e. rapid approach to steady-state). Our results are robust for a range of biologically reasonable parameters (we explored a range of morphogen production rates ($J=0.1-3{\mu}m^{-2}s^{-1}$), decay lengths ($\lambda=70-120{\mu}m$) and threshold positions ($x_T=150-250{\mu}m$)). In summary, pre-steady-state measurement does not provide increased robustness to internal fluctuations and, indeed, it can result in significantly increased positional error when the morphogen is a direct transcription factor, due to the small measuring volumes involved.

Rather than the absolute morphogen concentration used above, an embryo could alternatively interpret a normalised concentration \cite{Gurdon2001,Gregor2007b}. For example, the nuclear Bcd concentration appears to be approximately constant between cycles 9-14, although the extra-nuclear Bcd concentration may vary \cite{Gregor2007a}. In such systems, the numerator in (\ref{eq:omega0}) would be constant (here taken to equal the steady-state value) with changes in precision determined by the slope of the morphogen gradient at the threshold position. At early times the gradient is relatively flat at typical threshold distances, resulting in large positional error. The positional error decreases with time as the gradient becomes steeper at the threshold position. We also note that, compared with fluctuations in steady-state normalised concentrations, the effects of external and TAW fluctuations (see below) are increased by interpreting normalised morphogen concentrations in pre-steady-state. Therefore, considering normalised concentrations does not alter our conclusions.

\subsection{TAW Fluctuations}
Variations in the TAW affect the spatial precision of pre-steady-state morphogen gradients. The effect of different averaging onset times is illustrated for both the SDD and MRG models in Figure~2B. We denote the positional error due to such fluctuations ($\delta\tau$) by $\omega_{TAW}$. Fluctuations in $\tau_m$ can also alter $\omega_{TAW}$ (and $\omega_{int}$ due to an altered averaging period). However, in the regime $\tau_m\ll\tau$, such fluctuations make little difference to our results (data not shown) and are thus neglected. In this regime, the positional error due to fluctuations in the averaging onset time at position $x_T$ is (for $\delta{\tau}\ll\tau$)
\begin{equation}
\label{eq:time_RD}
\omega^i_{TAW} \approx g_i\lambda_{i}\left(\delta{\tau}/\tau\right) \,,
\end{equation}
where $g_{sdd}=\tau\partial_tf(x_T,\tau)[1-f(x_T,\tau)+\lambda_{sdd}\partial_xf(x_T,\tau)]^{-1}$ and $g_{mrg}=\nu\tau(e^{\nu\tau}-1)^{-1}$ (see Appendix B). Interestingly, we see that $g_{mrg}$ is independent of the threshold position and of $\lambda_{mrg}$, whereas $g_{sdd}$ is sensitive to all the kinetic parameters and the threshold position. The positional independence of $\omega^{mrg}_{TAW}$ arises because the concentration profile in the MRG model is always exponential, though with a time-dependent amplitude. Since morphogens typically control more than one target gene, $\omega^{mrg}_{TAW}$ being independent of position could be a possible advantage for the MRG model in pre-steady-state.

We now investigate the behaviour of $\omega_{TAW}$ numerically (see Appendix A). We first analyse how the positional error depends on the magnitude of the TAW fluctuations. An important result, clear from Figure~2C, is that the MRG model is typically more robust to TAW fluctuations than the SDD model. This result is general if both models have similar kinetic parameters, such as the decay length and degradation rate. We see that TAW fluctuations can be a significant source of positional error in pre-steady-state, even at small values of $\delta\tau/\tau$. However, the positional error not only depends on the magnitude of the TAW fluctuations, but also on the averaging onset time ($\tau$) itself, with the error being larger at early times. This is demonstrated in Figures~2D and~2E for the SDD and MRG models respectively, using $\delta\tau/\tau=10\%$.  Furthermore, we see in Figures~2C-E, that (\ref{eq:time_RD}) agrees very well with our numerical results.

We can draw three conclusions from the above results. First, pre-steady-state morphogen interpretation inevitably leads to an increase in measurement error due to TAW fluctuations. Given the high precision with which gene expression boundaries are typically defined \cite{Gregor2007b}, TAW fluctuations could be a significant impediment to pre-steady-state measurement. Note that this source of error is completely absent in steady-state where $g_i=0$ for both models.  Second, if a system has significant TAW fluctuations, the MRG model provides more precise positional information than the SDD model. Third, systems that interpret morphogen gradients in pre-steady-state must have additional regulatory mechanisms to ensure that fluctuations in the averaging onset time are well-controlled.

\subsection{External Fluctuations}

We now focus on the positional error due to external morphogen production fluctuations. The positional error due to such fluctuations, $\omega_{ext}$, has been studied previously \cite{Bergmann2007} and here we outline the results relevant to our conclusions. The positional error is (see Appendix B for details)
\begin{equation}
\label{eq:dJ}
\omega^i_{ext} \approx h_i\lambda_i(\delta{J}/J)
\end{equation}
where $h_{sdd}=[1+\lambda\partial_xf(x_T,\tau)(1-f(x_T,\tau))^{-1}]^{-1}$ and $h_{mrg}=1$ (and $\delta{J}/J=\delta{j_0}/j_0$ for MRG model). In steady-state $h_{sdd}=1$ and the two models have equal error. As with TAW fluctuations, $\omega_{ext}^{mrg}$ is independent of the threshold position whereas $\omega_{ext}^{sdd}$ is sensitive to the threshold position and the kinetic parameters. Since $\omega^{mrg}_{ext}$ is independent of position, MRG-like models could be favoured in pre-steady-state when multiple threshold boundaries are required.

We first investigate numerically how the positional error due to external production variations depends on when measurement occurs (see Appendix A). In Figure~2D we demonstrate that, for the SDD model with $\delta{J}/J=0.1$, the positional error is reduced by pre-steady-state measurement in agreement with \cite{Bergmann2007}. This increased robustness to such fluctuations is a major advantage of pre-steady-state measurement in the SDD model.  In comparison, Figure~2E shows that the MRG model (with $\delta{j}_0/j_0=0.1$) has constant positional error regardless of when measurement occurs. Moreover, the SDD model is typically more robust than the MRG model to source fluctuations, regardless of the size of the morphogen production fluctuations, since $h_{sdd}\le{h}_{mrg}$ (with $\lambda_{sdd}=\lambda_{mrg}$), see Figure~2F. Importantly, in all cases, we see excellent agreement between (\ref{eq:dJ}) and Figures~2D-F. Clearly, embryos with very large external production fluctuations will favour diffusive morphogen propagation with pre-steady-state measurement.

Morphogen gradients described by the SDD model benefit from pre-steady-state decoding when there are large external morphogen production fluctuations. However, such systems have increased sensitivity to TAW variations. Therefore, even in systems without significant internal fluctuations, it may still not be advantageous to interpret morphogens prior to steady-state.  Furthermore, in systems with considerable internal fluctuations, the SDD model can be particularly imprecise, discouraging pre-steady-state morphogen interpretation regardless of the external fluctuations. The MRG model is not favoured by morphogens with large external fluctuations. However, the MRG model displays increased robustness to TAW fluctuations and it is less sensitive to internal fluctuations at early times. If a developmental system has to interpret morphogens that are direct transcription factors prior to steady-state (the embryo needs a position quickly but not precisely) then an MRG-like model will likely be more effective than an SDD model so long as external fluctuations are not the dominant source of error.

\section{Experimental Application}

Having built-up a clear understanding of the precision of pre-steady-state morphogen gradients in the presence of fluctuations, we can now relate our theoretical approach to experiment. We consider two general developmental systems: first, when the morphogen is a direct transcription factor, such as Bcd; and second, morphogens interpreted by cell-surface receptors, such as Nodal. We account for TAW variations and internal and external fluctuations simultaneously to calculate the total positional error, $\epsilon$ (details in Appendix A). The precision given by $\epsilon$ is a lower bound on the error in determining $x_T$. We have neglected other sources of error, such as noise from transcription and translation of target genes \cite{Tkacik2008b} and cell-to-cell variability \cite{Lander2009}.  Furthermore, we assume the measurement process is perfect. Recent experiments on the Bcd-{\it hunchback (hb)} signaling pathway in {\it Drosophila} suggests that diffusive input noise dominates over transcriptional/translational noise \cite{Tkacik2008b}, supporting our approach. 

\subsection{Morphogen Transcription Factors}

Morphogens that are transcription factors are measured at nanometre-scale DNA binding sites (${\Delta}x\approx3nm$), so, as mentioned above, internal fluctuations can be significant sources of error \cite{He2008,Gregor2007b}. We focus on the Bcd-{\it hb} signaling pathway in {\it Drosophila} where {\it hb} is a target gene of Bcd \cite{Bergmann2007, Gregor2007a, Gregor2007b}. This network provides precise positional information, with the {\it hb} expression domain boundary defined to within 2\% of embryo length (EL) \cite{Houchmandzadeh2002,He2008,Manu2009,Gregor2007b}. Using the known kinetic parameters for Bcd \cite{Gregor2007a,Gregor2007b}, early measurement results in large positional error, primarily due to internal fluctuations, Figure~3A. This suggests that early Bcd interpretation (as soon as cycle 9, as proposed by \cite{Bergmann2007,Cell-Argument2008a}) could well be disfavoured. The experimentally observed diffusion constant for Bcd ($D=0.3\mu{m}^2s^{-1}$ \cite{Gregor2007a}) and the predicted value from Bergmann et. al ($D=1{\mu}m^2s^{-1}$ \cite{Bergmann2007}) are both insufficient to allow accurate positioning at $x_T\approx240{\mu}m$ after $90$ minutes, see Figure~3B (for each $D$, $J$ and $\mu$ are chosen to leave the steady-state Bcd profile unchanged). Our results include spatial averaging \cite{Gregor2007b} (see \cite{Erdmann2009,Okabe2009} for possible biochemical mechanisms for spatial averaging), without which internal fluctuations further penalise pre-steady-state decoding. In conclusion, Bcd does not appear well-suited to early interpretation.

However, Bcd does not independently determine the threshold position of its target genes. In the early {\it Drosophila} embryo a gap-gene network regulates positioning \cite{Jaeger2004b}. We analysed a simplified model of the system (see \cite{Bergmann2007,Jaeger2004a} and Appendix C) to check that our conclusions for the SDD model hold for more complex networks. In this model, a pre-steady-state Bcd gradient defines the positions of the gap genes, and these spatial positions are stabilised by gap gene interactions \cite{Bergmann2007}. Importantly, once a threshold position is defined, it does not alter significantly even though the morphogen concentration continues to increase. In Figure~3C we show the Hb concentration profiles from 20 simulations with fluctuations in $J$ only (red lines) and all three considered types of fluctuations (green lines). Figure~3C clearly shows that only considering external morphogen production fluctuations (with corresponding positional error $\epsilon_{sdd}\approx 2\%$EL) ignores significant sources of positional error ($\epsilon_{sdd}\approx6\%$EL including all fluctuations). Hence, even in this more complex model it seems unlikely that Bcd can provide accurate positional information after $90$ minutes (the timescale for {\it hb} expression) with $D=1{\mu}m^2s^{-1}$.

The above conclusions are deduced from a diffusive model of Bcd morphogen gradient formation. Recent experiments suggest that the {\it bcd} mRNA may not be localised in the anterior region of the embryo, but instead the {\it bcd} mRNA forms its own gradient across the embryo \cite{Spirov2009,Lipshitz2009}. The MRG model is a representation of such spatially distributed morphogen production. From Figure~3A, we find that for the MRG model pre-steady-state measurement does not grant improved performance over waiting for steady-state. In particular, the SDD and MRG models have similar positional error after 90 minutes. Therefore, we predict that even if Bcd is driven by an underlying mRNA gradient, pre-steady-state interpretation is not advantageous. The MRG model could perform better by increasing the rate of degradation, but the parameters are constrained by the known timescales for gradient formation and the Bcd steady-state profile. The MRG model can be used to describe Bcd gradient formation within the gap-gene system described above. Again, we find that the error in positioning of the {\it hb} expression domain boundary is too large when interpreting the MRG model prior to reaching steady-state (positional error of $\epsilon_{mrg}\approx3\%$EL when only $j_0$ varies but $\epsilon_{mrg}\approx 6\%$EL when we consider all three types of fluctuations).

\subsection{Morphogen Interpretation at a Cellular Level}

A second class of morphogens we consider are signaling molecules initially sampled at the cellular level by cell-surface receptors. In this case, there are two general stages to morphogen interpretation. First, the morphogen gradient must be interpreted precisely by cell surface receptors. Second, the receptor signal must be precisely decoded into a corresponding target gene response within the cell's nucleus.  Typically, for morphogens that are members of the transforming growth factor-$\beta$ family (such as Nodal and Activin \cite{Massague2005,Hill2001}), the binding of morphogen to a cell-receptor results in the phosphorylation of a member of the Smad-family of transcription factors \cite{Massague2005,Hill2001}. Stable phosphorylation of the appropriate Smad protein is achieved within 15-30 minutes of exposure to the morphogen-bound receptor \cite{Massague2005}. However, the corresponding timescale for target gene expression (for example, {\it Xbrachury} \cite{Gurdon1995}) is typically about an hour. 

Activin passively diffuses through the embryo over a scale of $300{\mu}m$ in a period of 5 hours \cite{Gurdon1994} ($D\approx2.5{\mu}m^2s^{-1}$) and its concentration profile in steady-state is exponentially decaying \cite{McDowell1997}; the SDD model should therefore provide a good picture. From the above discussion, we see that interpretation of Activin by cell surface receptors occurs relatively quickly, therefore the condition $\tau_m\ll\tau$ should approximately hold.  From Figure~3D, we see that precise pre-steady-state decoding is possible in such a scenario due to the effects of internal fluctuations on Activin interpretation being small (since the effective cell-wide measuring volume is large). The predicted precision is similar to that found in steady-state. Note that the interpretation of Activin is not an on-off reading (as occurs in the Bcd-{\em hb} system and as assumed by our calculation). Rather one gradient is transformed into another. Nevertheless our qualitative conclusions, namely that internal fluctuations are unimportant, will be unaffected by this distinction. Recent experiments on the Nodal signaling pathway in zebrafish have also demonstrated clear pre-steady-state behaviour, with the Nodal-induced gene expression of {\it no-tail} spreading with time to over $200{\mu}m$ in length \cite{Harvey2009}. Since Nodal is interpreted by cell-surface receptors, we predict that the effects of internal fluctuations are unlikely to be a significant source of error in {\it no-tail} boundary precision. Finally, Figure~3E demonstrates that the MRG model can also provide precise positional information for morphogens interpreted at a cellular level. 

We now consider the downstream expression of target genes. Using (\ref{eq:omega0}) and typical parameters estimated from the Bcd transcription factor, we find that an averaging time of about an hour is required for the downstream transcription factors to reduce the positional error to less than the width of the cell itself (assuming there is no spatial averaging). This calculation is consistent with the hour timescale for target gene expression.

We conclude that pre-steady-state decoding of morphogens interpreted by cell-surface receptors can occur. The difficulties posed by large internal fluctuations can be avoided by (i) sampling the morphogen at a cellular level, thereby ensuring a large measurement volume and (ii) using a long averaging time to reduce fluctuations in the downstream transcription factor.

\section{Discussion and Conclusions}

We have calculated the precision of positional information specified by pre-steady-state morphogen gradients due to internal and external fluctuations and variations in the onset time of measurement. Our results allow us to make the following general conclusions. First, morphogens that are direct transcription factors, and which are interpreted prior to steady-state, face considerable drawbacks due to large internal fluctuations. Second, even in systems without significant internal fluctuations (such as morphogens interpreted by cell-surface receptors), pre-steady-state interpretation is not necessarily favoured. The properties that confer increased robustness to external morphogen production fluctuations (small $D$, early measurement) are precisely the properties that increase sensitivity to TAW fluctuations (since the system is further from steady-state). So, when does it pay for a developmental system to `rush' morphogen interpretation? We disagree with Bergmann et al. \cite{Bergmann2007} that systems interpret morphogens in pre-steady-state to increase robustness to fluctuations. Rather, pre-steady-state precision is either reduced or is roughly the same as in steady-state. Therefore, alternative explanations for pre-steady-state decoding, such as the benefit of defining multiple genes at similar spatial positions \cite{Gurdon1995} or the need for rapid (but potentially imprecise) gene boundaries are more consistent with our analysis.

In this study, we focused on three-dimensional systems. In this case, larger detector regions do reduce the effects of internal noise (see Ref.~\cite{Tostevin2007}). Conversely, strictly in two-dimensions the size of the detector region has little effect on positional precision \cite{Tostevin2007}. For morphogen interpretation to occur effectively in two-dimensions, the depth of the system needs to be comparable to the size of the morphogen protein (i.e. a few nanometres). Even in relatively flat systems such as eye and wing imaginal disc patterning in {\it Drosophila} \cite{Wehrli1998, Lander2001, Teleman2004, Kicheva2007}, morphogen interpretation therefore effectively occurs in three-dimensions. Indeed, experiments on flattened {\it Xenopus} animal caps (only a few cells deep), find no appreciable effects on the precision of the Activin gradient \cite{Kinoshita2006}. However, it would be interesting to investigate the positional precision of morphogen gradients constrained to membranes as two-dimensional effects may then be more pronounced.

In our analysis, we also assumed that the morphogen gradient was sampled continuously in space across the system. However, in real systems the morphogen gradient is sampled discretely at nuclei or cell positions. Moreover, in the Bcd-{\it hb} system in {\it Drosophila}, the number of nuclei changes during the relevant developmental period, doubling at the end of each cycle. For this reason, it is revealing to scale the positional error by the internuclear distance, to see, for example, whether neighbouring nuclei can be distinguished. Since the internuclear distance during cycle 13-14 is around $2-3\%$EL, we see from Figure~\ref{fig3}A that in pre-steady-state it is never possible to distinguish neighbouring nuclei at mid-embryo using the Bcd gradient.

We have applied our results to several experimental systems. In particular, we focused on the Bcd-{\it hb} system in {\it Drosophila} which has provoked controversy regarding whether Bcd is interpreted prior to steady-state \cite{Cell-Argument2008a,Cell-Argument2008b}. If Bcd propagation is described by the SDD model (as proposed by \cite{Bergmann2007,Cell-Argument2008a}), we find that pre-steady-state interpretation results in increased positional error given the measured Bcd diffusion constant, even when we include the effects of spatial averaging. We have also seen that our conclusions hold when we include more complex dynamics, such as the gap-gene regulatory network. Similar results hold for the MRG model, with early measurement of Bcd disadvantaged by internal fluctuations. Hence, we predict that Bicoid is unlikely to be interpreted prior to reaching steady-state.  Activin and Nodal are, however, both consistent with our predictions for morphogens that can be interpreted prior to reaching steady-state. Since the relative size of a cell is large, cells can interpret morphogen gradients quickly as the effects of internal fluctuations are small. Moreover, the cell then uses longer averaging times for the downstream transcription factors to reduce the effects of internal fluctuations on target gene expression. Finally, we predict that systems employing pre-steady-state measurement may have additional regulatory mechanisms to ensure that fluctuations in the onset of averaging are reduced.

How morphogen gradients provide such precise spatial information has long been puzzling. Whilst pre-steady-state decoding of morphogen gradients does confer increased robustness to some fluctuations, it is certainly not a `cost-free' solution for developing embryos, particularly in the presence of significant internal fluctuations. Furthermore, we have demonstrated that fluctuations in the onset time of averaging can have as large an effect on precision as other sources of noise. By providing a quantitative approach to calculating the positional error due to multiple sources of fluctuations, we have outlined a framework for future experimental and theoretical studies of pre-steady-state morphogen gradients.

\subsection*{Acknowledgments}
We thank Richard Morris, Scott Grandison and Andrew Angel for helpful comments on the manuscript.

\appendix

\section{Calculating Positional Error} 
We find the concentration over a single realisation numerically,
\begin{equation}
\label{eq:av-sum}
{\rho}_{\tau,\tau_m}(x) = \frac{1}{N}\sum_{n=1}^N {\rho}(x,t_n)\,,
\end{equation}
where ${\rho}_{\tau,\tau_m}$ denotes the average morphogen concentration profile over a period $\tau_m$ starting at time $\tau$ after initiation, with $t_n=\delta{t}+t_{n-1}$ ($t_1=\tau$), $\delta{t}=\tau_{ind}$ and $N=\tau_m/\delta{t}$. For a particular gene expression boundary $x_T$, we first find the corresponding threshold concentration (in the absence of fluctuations) $\rho_T={\rho}_{\tau,\tau_m}(x_T)$. For each simulation $J$, $j_0$ and $\tau$ are drawn from Gaussian distributions with standard deviations $\delta{J}$, $\delta{j}_0$ and $\delta\tau$ respectively. (\ref{eq:av-sum}) is then used to calculate the concentration profile (without internal fluctuations) averaged over a period $\tau_m$ for each parameter realisation, with the error determined from finding where this profile goes through $\rho_T$. The positional error due to TAW and external fluctuations is found from 50 repetitions of the simulations to build the distribution that describes the error in determining $x_T$. In Figure~2, we only investigate TAW fluctuations or source fluctuations in each subplot (so, for example, in Figure~2C $\delta{J}=\delta{j_0}=0$).

To calculate the positional error due to internal fluctuations for each given parameter realisation we use (\ref{eq:omega0}). The constant $A=k_{\tau}/\sqrt{\Delta{x}D_0N_{spat}}$, where $N_{spat}$ denotes possible spatial averaging \cite{Gregor2007b} and $k_{\tau}$ is a system-dependent constant associated with time averaging (see \cite{Tostevin2007} for details). When calculating the total positional error, we simultaneously calculate the positional error from internal fluctuations $\omega_{int}$ using (\ref{eq:omega0}) and the error from TAW and source fluctuations, $\omega_{TAW,ext}$.  We find the total positional error due to internal noise together with TAW and morphogen production fluctuations using $\epsilon\approx(\omega_{TAW,ext}^2+\omega_{int}^2)^{1/2}$ (approximation valid if $\tau_m\ll\tau$). For all results presented, the mean measurement error and standard deviation were found from 10 independent computations of the positional error. When $\tau_{ind}$ is very small, $N$ becomes very large and we approximate the sums in (\ref{eq:omega0}) and~(\ref{eq:av-sum}) by integrals.

\section{Derivation of $\omega_{TAW}$ and $\omega_{ext}$} 
$\omega_{TAW}$: For small fluctuations in $\tau$, we can Taylor expand the concentration at $x_T$ to first order with respect to fluctuations in time, $\delta\rho \approx |\partial_t\rho| \delta{\tau}$ where $\delta\rho$ is the fluctuation away from the mean concentration value at position $x$ and time $\tau$. To leading order, $\omega \approx |1/\partial_{\rho}x| \delta\rho$ \cite{Tostevin2007} and hence $\omega_{TAW} \approx |\partial_t\rho/\partial_x\rho|\delta{\tau}$. Substituting in (\ref{eq:Solution1}) and~(\ref{eq:Model2}) we obtain (\ref{eq:time_RD}) and $g_{sdd}$ and $g_{MRG}$. $\omega_{ext}$: Similar to above, we Taylor expand the morphogen concentration with respect to fluctuations in $J$ (for SDD model) or $j$ (for MRG model) to find $\omega_{ext} \approx |1/\partial_{\rho}x| \delta\rho\approx|\partial_J\rho/\partial_x\rho|\delta{J}$.

\section{Gap-gene model}
The results presented in Figure~3C are based on the simplified reaction-diffusion model of the {\it Drosophila} gap gene regulatory network described in \cite{Bergmann2007}. The model incorporates the gap genes {\it giant, knirps} and {\it Kr$\ddot{u}$ppel} and their mutual repression of each other and {\it hb} \cite{Jaeger2004a,Jaeger2004b}. The Bcd morphogen gradient evolves with time, allowing an investigation of how the system responds to a pre-steady-state morphogen gradient. The gap genes and {\it hb} are activated by the Bcd gradient at different threshold concentrations. The gap genes are activated, on average, 90 minutes after Bcd is initiated, consistent with experiment \cite{Bergmann2007}. Repression, activation and autoactivation (gap genes promoting their own expression) are modelled using Hill functions. The diffusion rates for the products of the gap genes and {\it hb} are significantly less than the Bcd diffusion rate (i.e. they are effectively localised) but their production and degradation rates are significantly larger than the respective rates for Bcd.  Full details of equations and parameters used can be found in \cite{Bergmann2007}, though we use $\lambda_{sdd}=100{\mu}m$. In the simulations we alter $\tau$ by fluctuating the gap gene activation time. We also alter $\tau_m$, but since $\tau_m\ll\tau$ such fluctuations are not a major contributor to positional error. External fluctuations are incorporated by fluctuating the Bcd production rate, $J$. For each set of parameters, the threshold position is defined as where the normalised Hb concentration equals $0.5$ after time $\tau+\tau_m$. 

Whilst incorporating fluctuations in $J$, $\tau$ and $\tau_m$ is straightforward, dealing with stochastic fluctuations in the concentrations is more difficult. One solution would be to solve the set of differential equations stochastically, a complicated procedure. Instead, we fluctuate the threshold concentrations for each protein with deviation $A\tau_m^{-1/2}\rho_T^{1/2}$ and $\rho(x_T,\tau)=\rho_T$. Effectively, instead of having fixed threshold concentrations and a fluctuating morphogen gradient, we have a smooth morphogen gradient but fluctuating threshold concentrations. To demonstrate that this approach is reasonable we consider the error due to internal noise in steady-state using the SDD model.  In the former case, using (\ref{eq:omega0}), $\omega_{int}\approx A\tau_m^{-1/2}\langle\rho(x_T)\rangle^{1/2}/|\langle\partial_x\rho(x_T)\rangle|=A\lambda/\sqrt{\tau_m\rho_T}$ whereas for the latter scenario, $\omega_{int}\approx\lambda\delta\rho_T/\rho_T$ (for $\delta\rho_T\ll\rho_T$). Therefore, if $\delta\rho_T=A\tau_m^{-1/2}\rho_T^{1/2}\ll\rho_T$ the two approaches give the same positional error. This result extends straightforwardly to pre-steady-state measurement.

\noappendix

\pagebreak
\begin{figure}[!ht]
\begin{center}
   \includegraphics[width=4in]{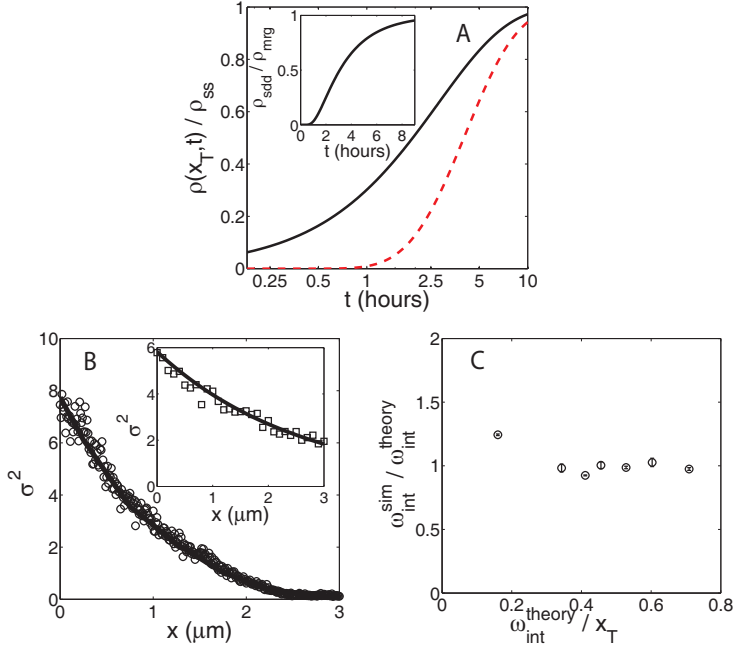}
\end{center}
\caption{
{\bf Pre-Steady-State Profiles and Fluctuations.}
{\bf A}: Average morphogen concentration (scaled by steady-state value) at fixed position $x_T$ against time after initiation (on log scale). Black line corresponds to the MRG model whilst dashed red line denotes the SDD model. $\rho_0=\rho_1=30{\mu}m^{-3}$, $D=1{\mu}m^2s^{-1}$, $\mu=\nu=10^{-4}s^{-1}$, $x_T=240{\mu}m$, $\lambda_{sdd}=\lambda_{mrg}=100{\mu}m$. Inset shows ratio of the concentrations in the SDD and MRG models at $x=x_T$ for varying times after initiation.
{\bf B}: Particle number variance, $\sigma^2$, from simulations of SDD model ($\circ$) with line denoting $\sigma^2=\langle{n}(x)\rangle$, the mean particle number, the prediction from Poisson statistics. Inset is same but for MRG model. $\lambda_{sdd}=\lambda_{mrg}=2.6{\mu}m$, $J=1{\mu}m^{-2}s^{-1}$, $j_0=0.67{\mu}m^{-3}s^{-1}$, $D=D_0=0.67{\mu}m^2s^{-1}$, $\mu=\nu=0.1s^{-1}$, $\tau=5s$, $\tau_m=1s$, $\Delta{x}=10^{-2}\mu{m}$.
{\bf C} Positional error $\omega_{int}$ from stochastic simulations \cite{Tostevin2007} compared with theoretical prediction for SDD model (\ref{eq:omega0}). Results for range of parameters: $5\le\tau\le60$ seconds, $1\le\tau_m\le20$ seconds, $2<x_T<7\mu{m}$, $D=D_0=0.67{\mu}m^2s^{-1}$, $0.01\le\mu\le0.1s^{-1}$, $1\le{J}\le10\mu{m}^{-2}s^{-1}$.
All simulations are three-dimensional, with $L_x=20{\mu}m$ and $L_y=L_z=3\mu{m}$.}
\label{fig1}
\end{figure}

\begin{figure}[!ht]
\begin{center}
\includegraphics[width=4in]{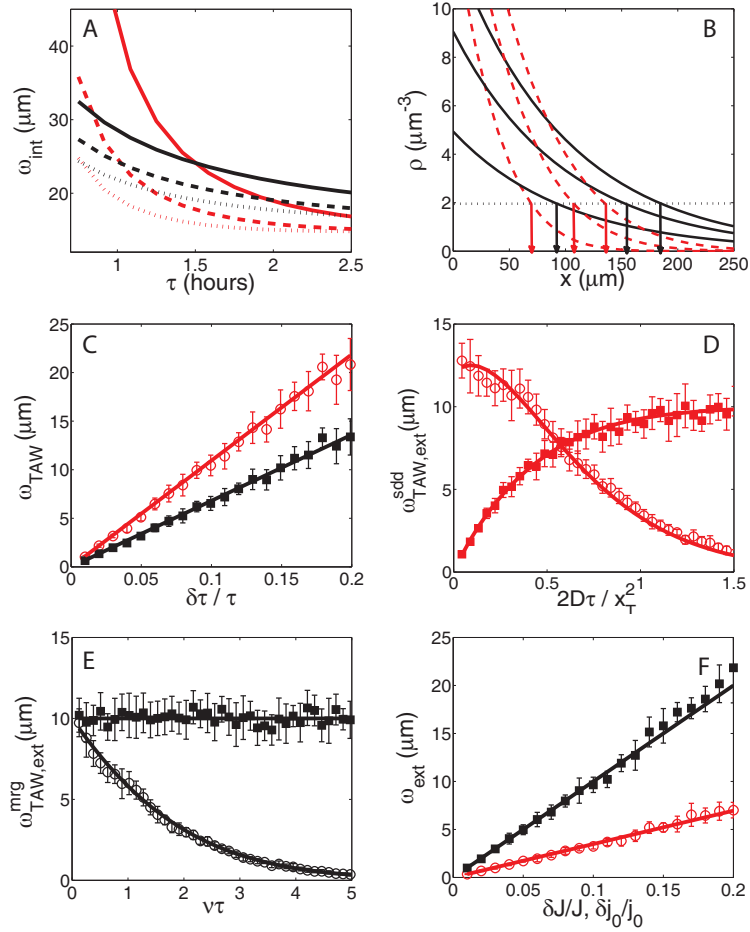}
\end{center}
\caption{
{\bf Positional error due to fluctuations.} (MRG Model: black lines and symbols. SDD Model: red lines and symbols).
{\bf A}: Positional error due to internal fluctuations, ${\omega}_{int}$, against averaging initiation time $\tau$. Solid, dashed and dotted lines correspond to $D=1,1.5,2{\mu}m^2s^{-1}$ for the SDD model and $\nu=1,1.5,2\times10^{-4}s^{-1}$ for the MRG model respectively. $J$, $j_0$ and $\mu$ are also altered such that each simulation has the same steady-state profile with $\rho_0=\rho_1=30{\mu}m^{-3}$, $\lambda=100{\mu}m$. 
{\bf B}: Morphogen concentrations for SDD and MRG models plotted at 30 minute intervals ($\tau=30$, $60$, $90$ minutes) with a fixed threshold concentration $2{\mu}m^{-3}$ (dotted line). The arrows denote how the measured threshold position varies with time in the SDD and MRG models.
{\bf C}: Positional error due to fluctuations in $\tau$, $\omega_{TAW}$, as a function of $\delta\tau/\tau$.
{\bf D}: Comparison of positional error $\omega^{sdd}_{TAW}$ ($\circ$) and $\omega_{ext}^{sdd}$ (filled squares) against $2D\tau/x_T^2$ with $\lambda_{sdd}$ constant ($x_T^2/2D$ is the approximate time for the system to be in steady-state at position $x_T$). 
{\bf E}: Comparison of positional errors $\omega^{mrg}_{TAW}$ ($\circ$) and $\omega^{mrg}_{ext}$ (filled squares) against $\nu\tau$. For $\nu\tau\gtrsim2.3$, the concentration is greater than $90\%$ of the steady-state concentration.
{\bf F}: Positional error due to external morphogen production fluctuations ${\omega}_{ext}$ as function of the relative magnitude of the source fluctuations $\delta{J}/J$ (for SDD model) and $\delta{j}_0/j_0$ (for MRG model). 
In Figures~{\bf C}-{\bf F}, solid lines are theoretical fits from (\ref{eq:time_RD}) and~(\ref{eq:dJ}).
Parameters used, unless stated otherwise, $x_T=240{\mu}m$ \cite{Houchmandzadeh2002}, $\tau=85$ minutes \cite{Bergmann2007}, $\tau_m=5$ minutes, $L=500{\mu}m$ \cite{Houchmandzadeh2002}, $J=0.3{\mu}m^{-2}s^{-1}$, $j_0=3\times10^{-3}{\mu}m^{-3}s^{-1}$, $D=1{\mu}m^2s^{-1}$ \cite{Bergmann2007}, $\mu=\nu=10^{-4}s^{-1}$, $\lambda_{sdd}=\lambda_{mrg}=100{\mu}m$ \cite{Houchmandzadeh2002}, $k_{\tau}=0.56$ \cite{Saunders2009}, ${\Delta}x=3{\times}10^{-3}{\mu}m$ \cite{Gregor2007b}, $D_0=0.3{\mu}m^2s^{-1}$ \cite{Gregor2007a}, $N_{spat}=0.06\tau_m$ \cite{Gregor2007b}, $\delta{J}/J=0.1$, $\delta{j_0}/j_0=0.1$, $\delta\tau/\tau=0.1$.
}
\label{fig2} 
\end{figure}

\begin{figure}[!ht]
\begin{center}
\includegraphics[width=4in]{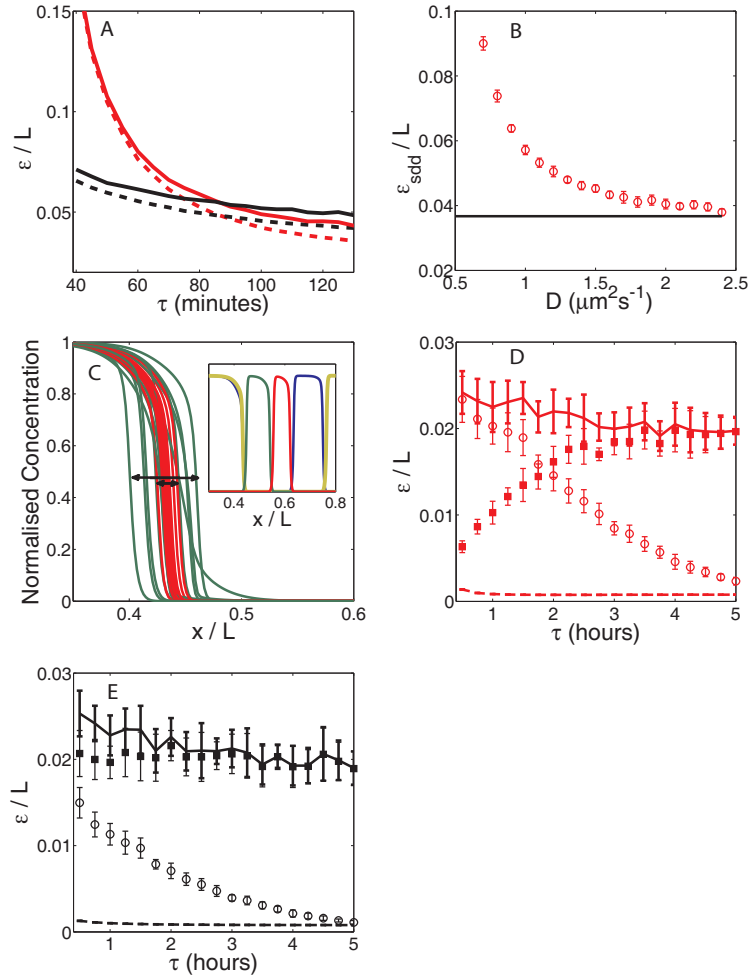}
\end{center}
\caption{ 
{\bf Total positional error.} (Same parameters and notation as Figure~2 unless stated otherwise).
{\bf A}: Total positional error scaled by system length $\epsilon/L$ (solid lines) against averaging initiation time, $\tau$. Dashed lines correspond to the positional error solely due to internal fluctuations (TAW and external fluctuations not shown for clarity).
{\bf B}: Total positional error in SDD model scaled by system length $\epsilon_{sdd}/L$ against rate of diffusion $D$ with $\tau=90$ minutes, keeping $\lambda$ and $\rho_0=J\lambda/D$ constant so that we compare with the same Bcd steady-state profile (line denotes the steady-state value of $\epsilon$).
{\bf C}: (Normalised) Hb concentration against relative embryo position in simplified gap-gene regulatory network model (see Appendix C and \cite{Bergmann2007} for details). Red lines correspond to fluctuations only in $J$ ($\delta{J}/J=0.1$). Green lines include all three sources of error considered (parameters for Bcd as {\bf A}). Arrows denote positional error when relative Hb concentration is 0.5. Inset: concentration profiles of Hb (yellow) and gap gene products \cite{Jaeger2004a} (red={\it knirps}, green={\it kruppel}, blue={\it giant}).
{\bf D}: Same as {\bf A} except ${\Delta}x=10{\mu}m$, $D=D_0=2.5{\mu}m^2s^{-1}$, $x_T=280{\mu}m$ \cite{Gurdon1994} and no spatial averaging, with $\fullsquare$ and $\circ$ denoting the positional error due to external and TAW fluctuations respectively in the SDD model.
{\bf E}: Same as {\bf D} except for MRG model with $\nu=2.5\times10^{-4}s^{-1}$.
}
\label{fig3}
\end{figure}

\end{document}